\documentstyle[11pt]{article}

\begin{document}

\begin{center}

{\Large {\bf Minimizing Effective Many-Body Interactions}}

\vspace{0.2in}

B. R. Barrett$^{1)}$, D. C. Zheng$^{1)}$,
R. J. McCarthy$^{2)}$ and J. P. Vary$^{3,4)}$

\end{center}

\vspace{0.2in}
\begin{small}

\hspace{-0.3in}
$^{1)}$
{\it Department of Physics, University of Arizona, Tucson, Arizona 85721}

\hspace{-0.3in}
$^{2)}$
{\it Department of Physics, Kent State University, Asatabula, Ohio 44004}

\hspace{-0.3in}
$^{3)}$
{\it Department of Physics and Astronomy, Iowa State University, Ames,
					IA 50011}

\hspace{-0.3in}
$^{4)}$
{\it Institute for Theoretical Physics, University of Heidelberg,
					Heidelberg, Germany}
\end{small}

\thispagestyle{empty}

\vspace{0.5in}

\begin{abstract}
A simple two-level model is developed and used to test the properties
of effective interactions for performing nuclear structure calculations
in truncated model spaces.
It is shown that the effective many-body interactions
sensitively depend on the choice of the single-particle basis and
they appear to be minimized when a self-consistent Hartree-Fock basis is used.
\end{abstract}

\pagebreak

\section{Introduction}
Assuming that only two-body interactions act among nucleons in nuclei,
we can write the nuclear Hamiltonian as
\begin{equation}
H = \sum_{i=1}^{A}\frac{\mbox{\boldmath $p$}_{i}^2}{2m} + \sum_{i<j}v_{ij}\; .
\end{equation}
This Hamiltonian can be divided into two terms,
$H_0$ and $H_I$, as
\begin{equation}
H = \sum_{i=1}^{A} \left(
\frac{\mbox{\boldmath $p$}_{i}^2}{2m} + u_i\right)
+ \left(\sum_{i<j}^{A} v_{ij} - \sum_{i=1}^{A} u_{i}\right)
\equiv H_0 + H_I,
\end{equation}
where $H_0$ is the one-body Hamiltonian which defines a single-particle
basis and $H_I$ is the two-body ``residual'' interaction.
If the Schr\"{o}dinger equation
\begin{equation}
H\Psi_i(1,2,\ldots,A) = E_i \Psi_i(1,2,\ldots,A)   \label{schr}
\end{equation}
could be solved in the infinite Hilbert space for a many-body system,
the results would not depend on the choice of the one-body potential
$u$, or, equivalently, the single-particle basis. In practice, in order
to solve Eq.(\ref{schr}), one must
truncate the infinite Hilbert space to a finite model space and
introduce an effective interaction ($V_{\rm eff}$) to
be used in the truncated model space.
For an $A$-nucleon system, the effective interaction
$V_{\rm eff}$ will, in principle, have two- [$V^{(2)}_{\rm eff}$],
three- [$V^{(3)}_{\rm eff}$], ..., and A-body [$V^{(A)}_{\rm eff}$] parts.
If the exact effective interaction, containing all (two- to $A$-body)
components, can be obtained, the results will again
be independent of the choice of the one-body potential $u$.
However, three- and more-body effective interactions are difficult
to calculate and are often ignored in practical shell-model calculations
with the hope that they are small. One is then left with only
the two-body effective interaction $V^{(2)}_{\rm eff}$ for performing
nuclear structure calculations in a severely truncated model space.

With recent developments on the effective interaction theory,
i.e., the use of the no-core model space \cite{nc1,nc2}
and the Lee-Suzuki approach \cite{ls,pb,zvb} to
the folded-diagram series, we can now calculate the effective
two-body interaction accurately.
The question is how to choose $u$ to give the best approximation
to the exact results when retaining only the $V^{(2)}_{\rm eff}$ part
of the effective Hamiltonian.

In this work, we will use an exactly soluble two-level model (so that
exact eigenenergies can be obtained) to show
that the contribution to the ground-state energy
from the often-neglected many-body effective interactions
depend quite sensitively on the choice of the one-body potential $u$.
We show that the Hartree-Fock (HF) self-consistent one-body field,
appears to minimize the three- and more-body effects and consequently,
the two-body effective interaction alone becomes a good approximation.

This paper is organized as follows. In the next section, we give a brief
review of the effective interaction theories.
In section 3, we introduce a simple
two-level model and calculate the energy-independent two-body effective
interaction for an arbitrarily chosen,
non self-consistent, single-particle basis.
In section 4, we repeat the calculation performed in section 3 for a
self-consistent Hartree-Fock basis and compare the results with those
obtained in section 3. Finally in section 5, we give our conclusions.

\section{Energy-Independent Effective Interactions}
When the full Hilbert space is divided into $P$ and $Q$ spaces with $P$
projecting out the model space from the full Hilbert space
and $Q$ projecting out the excluded space,
the $P$-space effective interaction for an eigenstate with an eigenenergy
$E$ can be written \cite{feshbach}
\begin{equation}
V_{\rm eff}(E) = PH_IP + PH_IQ \frac{1}{E - QHQ} QH_IP \; , \label{vAeff}
\end{equation}
Note that $E$, the energy of the many-body system which we wish to calculate,
appears in the right-hand-side of the above equation. For this reason, the
above effective interaction is also referred to as the energy-dependent
effective interaction.

About ten years ago,
Lee and Suzuki \cite{ls} proposed a method for calculating an
energy-independent effective interaction. This method is based on
a ``generalized'' G matrix, which is defined as
\begin{equation}
G(\omega) = PH_IP + PH_IQ \frac{1}{\omega - QHQ} QH_IP \; . \label{gA}
\end{equation}
This generalized G matrix, which depends on a starting energy $\omega$,
is often referred to as the Q-box \cite{qbox}.
Obviously, the generalized G matrix, or the Q-box, becomes
the energy-dependent effective interaction $V_{\rm eff}(E)$ when
$\omega$ is chosen to be equal to $E$.

It is easy to see that for a two-particle system ($A$=2),
the generalized G matrix becomes the Brueckner two-body G matrix \cite{bruc}.
Exact methods \cite{bhm,vy} exist for calculating the Brueckner G matrix
as a function of the starting energy $\omega$.
However, the generalized G matrix is an $A$-body operator
and is generally difficult to evaluate when $A>2$.
{\it If}\hspace{0.025in} it ever becomes possible to calculate
$G(\omega)$, the {\it energy-independent} effective interaction
$V_{\rm eff}$ can then be obtained
through an iterative procedure proposed in Ref.\cite{ls}.
Applying the Lee-Suzuki iterative procedure is equivalent to summing
over all the folded diagrams which are not included in the
generalized $G$ matrix calculation.
Further details on the Lee-Suzuki iterative
methods can be found in Refs.\cite{ls,pb,zvb}.

In the case of a one-dimensional model space (one can always choose the
model space to be one-dimensional no matter how many particles are
involved), the effective interaction $V_{\rm eff}$ for the model space
becomes a number (i.e., a one-by-one matrix).
It is then apparent by comparing Eq.(\ref{vAeff}) and Eq.(\ref{gA})
that when the starting energy $\omega$ happens to be one of the
eigenenergies of the system, $E$, one has
\begin{equation}
V_{\rm eff} = V_{\rm eff}(E) = G(\omega=E)\; ,    \label{vAgA}
\end{equation}
i.e., there is no need to utilize the iterative procedure and, therefore,
the contribution from the folded diagrams vanishes. In this
case, no iterations are needed to obtain the effective interaction
$V_{\rm eff}$ from $G(\omega)$.

In other words, in the case of a one-dimensional model space,
the Lee-Suzuki approach to $V_{\rm eff}$ simplifies to
the following equation for $E$:
\begin{equation}
E = PH_0P + G(E)\; ,		\label{ege}
\end{equation}
which may be solved graphically.
This has been shown explicitly in Ref.\cite{zvb}.
It is also shown in Ref.\cite{zvb} that
Eq.(\ref{ege}) is equivalent to the secular equation for the eigenenergies:
\begin{equation}
\det ( H - E I) = 0 \; .
\end{equation}
Therefore, the solutions to Eq.(\ref{ege}) correspond to the exact
eigenenergies of the $A$-body system.
Where it has been tested the method appears to work well.
Of course, it can only produce eigenvalues for states having non-vanishing
overlap with the model-space state.

Note that when the model space is $D$-dimensional ($D>1$),
the energy-dependent
effective interaction $V_{\rm eff}(E)$ is different from the
the LS energy-independent effective interaction $V_{\rm eff}$.
Both the model space Hamiltonians
$[H_0+V_{\rm eff}(E)]$ and $[H_0+V_{\rm eff}]$ have $D$ eigenvalues.
But for the former Hamiltonian, only one [$E$, i.e., the one at which
$V_{\rm eff}(E)$ is evaluated] out of $D$ eigenvalues
corresponds to a true eigenenergy of the system
while, for the latter Hamiltonian, all its $D$ eigenvalues are
true eigenenergies of the system.

Because it is {\it not} known how to compute $G(\omega)$ for
a system of $A$ nucleons, the usual procedure is to start with the
nuclear (or Brueckner) two-body reaction matrix $G^{(2)}$ \cite{bruc} and
try to calculate the two-body effective interaction $V^{(2)}_{\rm eff}$.
Assuming that the effective three- and higher-body forces are small,
one can then use such an effective two-body interaction as
input to shell-model programs, such as the {\sc oxbash} code \cite{oxbash}
to perform model-space diagonalizations for the entire
system. We briefly outline how one obtains the effective
two-body interaction $V^{(2)}_{\rm eff}$ from the starting-energy-dependent
Brueckner two-body G matrix.

The Brueckner G matrix represents the
infinite summation (ladder sum) of two-particle scatterings.
This is defined, in analogy to Eq.(\ref{gA}) for $G(\omega)$, as
\begin{equation}
G^{(2)}(\omega_2) = P_{2}H_IP_{2} + P_{2}H_I Q_{2}
	\frac{1}{\omega_2 -Q_{2}HQ_{2}} Q_{2} H_I P_{2}\; , \label{g2}
\end{equation}
where $P_{2}$ and $Q_{2}$ are now two-particle projection operators,
which determine the allowed and forbidden
intermediate states into which the two particles can scatter.
Note that $G^{(2)}$ is defined to depend on a starting energy
$\omega_2$. It is clear that this starting energy refers to a
two-particle system and when one approximates $V_{\rm eff}$
by omitting many-body interactions, $\omega_2$ is different from
the starting energy $\omega$ in $G(\omega)$.

The $\omega_2$-independent two-body effective interaction
$V^{(2)}_{\rm eff}$ can be obtained from $G^{(2)}(\omega_2)$
by summing over all the folded diagrams with two valence lines. This can now
be accomplished without much difficulty
by applying the Lee-Suzuki iterative procedure \cite{ls,pb,zvb}.

\section{$V^{(2)}_{\rm eff}$ for an Arbitrary Basis}
A simple two-level model, consisiting of two single-particle states
$|1\rangle$ and $|2\rangle$, is
used for our investigation. Each level can hold up to four nucleons:
spin-up proton ($p\uparrow$), spin-down proton ($p\downarrow$)
spin-up neutron ($n\uparrow$) and spin-down neutron ($n\downarrow$).
The Hamiltonian for this model is determined by $H_0$, represented by
the single-particle energies ({\sc spe})
of the two levels ($\epsilon_1$ and $\epsilon_2$),
and $H_I$, represented by 14 antisymmetrized, normalized
two-body matrix elements ({\sc tbme})
[$\langle ab| H_I|cd\rangle_{J,T}$].
Our choice for the {\sc spe} and the {\sc tbme} is given
in Table 1.

As our first calculation, we diagonalize, using the {\sc oxbash} shell-model
code \cite{oxbash}, the Hamiltonian
in the {\it full} two-level space for $A$=2, 3 and 4.
The results obtained are exact and are given in Table 2 in the column
under the heading ``Exact''.

\begin{table}
{\small \caption{The one-body part $H_0$, defined by the
{\sc spe} $\epsilon_1$ and $\epsilon_2$,
and the two-body part $H_I$, defined by the antisymmetrized,
normalized {\sc tbme}
$\langle ab|H_I|cd\rangle_{J,T}$, of the Hamiltonian $H$.
For later convenience, we also denote each matrix element
by $a_{ij}$, $b_{ij}$, {\it etc.}.}}
\begin{small}
\begin{center}
\begin{tabular}{l|l}\hline\hline
\multicolumn{2}{c}{$\epsilon_1=0 $, \hspace{1.0in}$\epsilon_2=10 $}\\ \hline
$a_{11}=\langle 11|H_I|11\rangle_{0,1}$=--3.0 &
$b_{11}=\langle 11|H_I|11\rangle_{1,0}$=--4.0 \\
$a_{12}=\langle 11|H_I|12\rangle_{0,1}$=--1.5 &
$b_{12}=\langle 11|H_I|12\rangle_{1,0}$=--2.0 \\
$a_{13}=\langle 11|H_I|22\rangle_{0,1}$=--1.7 &
$b_{13}=\langle 11|H_I|22\rangle_{1,0}$=--1.9 \\
$a_{22}=\langle 12|H_I|12\rangle_{0,1}$=--2.7 &
$b_{22}=\langle 12|H_I|12\rangle_{1,0}$=--2.9 \\
$a_{23}=\langle 12|H_I|22\rangle_{0,1}$=--1.6 &
$b_{23}=\langle 12|H_I|22\rangle_{1,0}$=--2.1   \\
$a_{33}=\langle 22|H_I|22\rangle_{0,1}$=--2.5 &
$b_{33}=\langle 22|H_I|22\rangle_{1,0}$=--2.9   \\
$c_{11}=\langle 12|H_I|12\rangle_{0,0}$=  0.0 &
$d_{11}=\langle 12|H_I|12\rangle_{1,1}$=  0.0 \\
					\hline\hline
\end{tabular}
\end{center}
\end{small}
\end{table}

Next we truncate the two-level space to a smaller space containing
only the lower level, which means that we must construct the effective
Hamiltonian appropriate for this model space.

As we mentioned in the Introduction, for the truncated space,
the effective interaction $V_{\rm eff}$ for an $A$-body system
can be written as
\begin{equation}
V_{\rm eff} = V^{(2)}_{\rm eff} + V^{(3)}_{\rm eff} + \cdots
+ V^{(A)}_{\rm eff} \; .		\label{veff}
\end{equation}
When only the two-body part
$V^{(2)}_{\rm eff}$ is kept in shell-model calculations,
the choice of the single-particle basis
becomes important as we shall see below.
In this section, we will use the original
single-particle basis ($|1\rangle$ and $|2\rangle$,
as used in Table 1),
which is not a self-consistent basis, to
calculate the effective two-body interaction $V^{(2)}_{\rm eff}$.
In the next section, we will introduce a self-consistent HF basis for
$A$=4 and re-calculate $V^{(2)}_{\rm eff}$ and compare
the results obtained in these two sections.

\begin{table}
{\small \caption{The $A$=2, 3, and 4 ground-state energies in the
two-level model from {\it exact} matrix diagonalizations and from the
one-level model space calculations with a two-body effective Hamiltonian
obtained in the arbitrary and in the self-consistent HF bases.}}

\begin{small}
\begin{center}
\begin{tabular}{l|rcc}\hline\hline
Ground-state energy    & Exact  & Arbitrary Basis & Self-Consistent Basis
								\\ \hline
$E_2$($J$=0, $T$=1)    & --3.390 & --3.390 & --3.390 \\
$E_2$($J$=1, $T$=0)    & --4.582 & --4.582 & --4.582 \\
$E_3$($J$=1/2, $T$=1/2)&--12.938 &--11.958 &  \\
$E_4$($J$=0, $T$=0)    &--27.703 &--23.916 &--27.645 \\ \hline\hline
\end{tabular}
\end{center}
\end{small}
\end{table}

To obtain the energy-independent effective two-body interaction
$V^{(2)}_{\rm eff}$, we
start with the starting energy-dependent two-body G matrix and
apply the Lee-Suzuki procedure \cite{ls}. Previously we labelled
this as $G^{(2)}$ but now we relabel it as $G^{JT}$ to indicate
the two-body conserved quantities.
For the Hamiltonian and single-particle basis
defined in Table 1, the $G^{JT}$ matrix element
for the $J$=0, $T$=1 channel for the chosen model space is given by
\begin{small}
\begin{eqnarray}
G^{01}(\omega_2) &=& a_{11}+
	\left( \begin{array}{cc} a_{12}, & a_{13}\end{array}\right)
 \left( \begin{array}{cc} \omega_2-(\epsilon_1+\epsilon_2+a_{22}) & -a_{23} \\
-a_{32} & \omega_2-(2\epsilon_2+a_{33}) \end{array}\right)^{-1}
	\left( \begin{array}{c} a_{21} \\ a_{31}\end{array}\right)
		 \nonumber \\
&=& (-3.0)+ \left( \begin{array}{cc} -1.5, & -1.7 \end{array}\right)
 \left( \begin{array}{cc} \omega_2-7.3 & 1.6 \\
	1.6 & \omega_2-17.5 \end{array}\right)^{-1}
\left( \begin{array}{c} -1.5 \\ -1.7 \end{array}\right)\; .  \label{g01}
\end{eqnarray}
\end{small}
Similarly, for the $J$=1, $T$=0 channel, we have
\begin{small}
\begin{equation}
G^{10}(\omega_2)= (-4.0)+
 \left( \begin{array}{cc} -2.0, & -1.9 \end{array}\right)
 \left( \begin{array}{cc} \omega_2-7.1 & 2.1 \\
	2.1 & \omega_2-17.1 \end{array}\right)^{-1}
\left( \begin{array}{c} -2.0 \\ -1.9 \end{array}\right)\; .  \label{g10}
\end{equation}
\end{small}

Clearly, these two-particle G matrix elements depend on the starting energy
$\omega_2$. As in Ref.\cite{pb,zvb},
we now use the Lee-Suzuki method \cite{ls} to obtain
the energy-independent two-body effective interaction, which is equivalent
to summing over the two-particle folded diagrams to all orders.
We obtain the following results:
\begin{equation}
(V^{(2)}_{\rm eff})^{01} = \left\{ \begin{array}{l}
-3.390 \hspace{0.3in} {\rm if}\; \omega_2<1.976\; , \\
\hspace{0.1in} 7.342 \hspace{0.3in} {\rm if}\;  1.976< \omega_2< 12.595\; , \\
17.848 \hspace{0.3in} {\rm if}\; \omega_2>12.595\; , \end{array}\right.
	\label{v01}
\end{equation}
and
\begin{equation}
(V^{(2)}_{\rm eff})^{10} = \left\{ \begin{array}{l}
-4.582 \hspace{0.3in} {\rm if}\; \omega_2< 1.288\; , \\
\hspace{0.1in} 7.157 \hspace{0.3in} {\rm if}\; 1.288 < \omega_2<12.391\; , \\
17.625 \hspace{0.3in} {\rm if}\; \omega_2>12.391\; . \end{array}\right.
	\label{v10}
\end{equation}
It is easy to verify that the above numbers
are precisely the eigenenergies of the two-particle system
($A$=2) whose Hamiltonian is defined by Table 1,
as they must be \cite{zvb},
since it is just an application of Eqs.(\ref{vAgA}) and (\ref{ege})
to a simple $A$=2 case.

The ground-state energies of the $A$=3 and $A$=4 systems, when
we neglect effective many-body terms in Eq.(\ref{veff}), can be expressed
in terms of these two matrix elements as
\begin{equation}
E_3 = 3\epsilon_1+
1.5\left[ (V^{(2)}_{\rm eff})^{01} + (V^{(2)}_{\rm eff})^{10}\right]\; ,
\end{equation}
and
\begin{equation}
E_4 = 4\epsilon_1+
3.0\left[ (V^{(2)}_{\rm eff})^{01} + (V^{(2)}_{\rm eff})^{10}\right]\; .
					\label{e4g}
\end{equation}
The results are (noting that $\epsilon_1$=0):
\begin{equation}
E_3 = 1.5(-3.390-4.582) = -11.958 \; ,
\end{equation}
and
\begin{equation}
E_4 = 3.0(-3.390-4.582) = -23.916 \; .
\end{equation}
These differ from the {\it exact} results of --12.938 for $E_3$
and --27.703 for $E_4$ by about 1.0 and 3.8, respectively.
The discrepances reflect the importance of the
neglected effective three- and (for $A$=4) four-body terms.

\section{$V^{(2)}_{\rm eff}$ for a Self-Consistent Basis}
Addressing the case with the largest discrepancy,
we now consider a self-consistent HF single-particle basis for the ground
state of the $A$=4 system. We rewrite the original Hamiltonian of
Table 1 as
\begin{equation}
H = (H_0+U) + (H_I-U) \equiv H'_0 + H'_I
\end{equation}
In the previous section, we have set $U$=0. In this section,
we use a self-consistent one-body field $U$ generated by all
the $A$ particles in the system. It can be obtained iteratively
by using the following equations where $n$ represents the iteration number:
\begin{eqnarray}
\langle \alpha_n |U_n| \beta_n\rangle
	&=&\sum_{J,T,\gamma={\rm occ.}}
	   \frac{(2J+1)(2T+1)}{2(2j_{\alpha}+1)}
           \sqrt{(1+\delta_{\alpha_{n-1},\gamma_{n-1}})
          (1+\delta_{\beta_{n-1},\gamma_{n-1}})}\nonumber \\
        & & \hspace{0.5in} \langle \alpha_{n-1}\gamma_{n-1}
	| H_I |\beta_{n-1} \gamma_{n-1}\rangle_{J,T} \; ,     \\
(H_0+U_n) |\alpha_n\rangle &=& \epsilon_{\alpha_n} |\alpha_n\rangle\; ,
\end{eqnarray}
where $|\alpha_i\rangle$, $|\beta_i\rangle$, $|\gamma_i\rangle$, {\it etc.}
are the single-particle states in the $i$-th iteration
and the summation ($\gamma$) is over the occupied states. The
{\sc tbme} in the above expression are normalized and antisymmetrized.
When converged, we obtain a self-consistent HF basis.

It is obvious from the above equations that the resulting
single-particle basis, defined by $H'_0=(H_0+U)$,
and the two-body residual interaction $H'_I=(H_I-U)$
are mass-dependent and
should only be applied to the system for which they are calculated.

It is easy to work out the
self-consistent HF basis for $A$=4, which is a ``closed-shell'' system
in our two-level model. The new single-particle
states, denoted by $|1'\rangle$ and $|2'\rangle$,
are linear combinations of the old ones:
\begin{eqnarray}
|1'\rangle &=& \hspace{0.15in}0.92785 |1\rangle + 0.37295 |2\rangle,  \\
|2'\rangle &=&-0.37295 |1\rangle + 0.92785 |2\rangle.
\end{eqnarray}
The corresponding single-particle energies are:
\begin{eqnarray}
H'_0|1'\rangle &=& -15.0323 |1'\rangle, \\
H'_0|2'\rangle &=& \hspace{0.2in}  5.3808 |2'\rangle .
\end{eqnarray}
The {\sc tbme} of the residual interaction ($H'_I=H_I-U$),
evaluated using the new basis, are given in Table 3.

Note that the {\sc tbme} of $H'_I$ listed in
Table 3 are $A$-dependent
not only because the one-body potential $U$ is $A$-dependent but also
due to the fact that when {\sc tbme} are calculated for
a one-body potential, a factor $1/(A-1)$ has to be introduced:
\begin{equation}
U = \sum_{i=1}^{A} u_i
  = \frac{1}{A-1}\sum_{i<j}^A(u_i+u_j).
\end{equation}

With the self-consistent single-particle basis, there is
no coupling between the $0p$-$0h$ configuration and the
$1p$-$1h$ configuration. This is guaranteed by the following
equation:
\begin{equation}
\sum_{J,T}(2J+1)(2T+1)\langle 1'1'|H'_I|1'2'\rangle_{J,T} = 0. \label{abcd}
\end{equation}
Now in the HF basis, we can obtain an estimate of
the ground-state energy of $A$=4 even before calculating $G^{JT}$
by using the lowest-order in $H'_I$ estimate of $V^{(2)}_{\rm eff}$
in Eq.(\ref{e4g}):
\begin{equation}
E_4^{\rm HF} = 4\epsilon_{1'} + 3\left( \langle 1'1'|H'_I|1'1'\rangle_{0,1}
	+ \langle 1'1'|H'_I|1'1'\rangle_{1,0}\right),
\end{equation}
which gives $E_4^{\rm HF}$=-27.283 MeV.
Note that this result is already
closer to the exact energy of -27.703 MeV
than the result obtained in the previous section for $A$=4.

\begin{table}
{\small \caption{The single-particle energies of $H'_0=(H_0+U)$ and
the two-body matrix elements of $H'_I=(H_I-U)$ for the Hamiltonian defined
in Table 1 with the self-consistent HF single-particle basis:
$|1'\rangle = 0.92785 |1\rangle + 0.37295 |2\rangle$, and
$|2'\rangle =-0.37295 |1\rangle + 0.92785 |2\rangle$.}}
\begin{small}
\begin{center}
\begin{tabular}{l|l}\hline\hline
\multicolumn{2}{c}{$\epsilon_{1'}=-15.0323 $, \hspace{0.6in}
	           $\epsilon_{2'}=  5.3808  $}\\ \hline
$\langle 1'1'|H'_I|1'1'\rangle_{0,1}$=  6.1415 &
$\langle 1'1'|H'_I|1'1'\rangle_{1,0}$=  4.8074 \\
$\langle 1'1'|H'_I|1'2'\rangle_{0,1}$=  0.0541 &
$\langle 1'1'|H'_I|1'2'\rangle_{1,0}$=--0.0541\\
$\langle 1'1'|H'_I|2'2'\rangle_{0,1}$=--1.3402 &
$\langle 1'1'|H'_I|2'2'\rangle_{1,0}$=--1.6120 \\
$\langle 1'2'|H'_I|1'2'\rangle_{0,1}$=  4.5702 &
$\langle 1'2'|H'_I|1'2'\rangle_{1,0}$=  4.2265 \\
$\langle 1'2'|H'_I|2'2'\rangle_{0,1}$=  1.2156 &
$\langle 1'2'|H'_I|2'2'\rangle_{1,0}$=  0.8955 \\
$\langle 2'2'|H'_I|2'2'\rangle_{0,1}$=  0.7399 &
$\langle 2'2'|H'_I|2'2'\rangle_{1,0}$=  0.8176 \\
$\langle 1'2'|H'_I|1'2'\rangle_{0,0}$=  6.5505 &
$\langle 1'2'|H'_I|1'2'\rangle_{1,1}$=  6.5505 \\  \hline\hline
\end{tabular}
\end{center}
\end{small}
\end{table}

For the HF basis, we then calculate
the effective two-body interaction for the truncated model space.
The results are
\begin{equation}
(V^{(2)}_{\rm eff})^{01} = 6.091\; , \hspace{0.5in}
(V^{(2)}_{\rm eff})^{10} = 4.738\; .
\end{equation}
So the ground-state energy for $A$=4,  using an effective two-body
interaction in a self-consistent basis, is
\begin{equation}
E_4 = 4\epsilon_{1'} + 3.0\left[
(V^{(2)}_{\rm eff})^{01} + (V^{(2)}_{\rm eff})^{10}\right]
= -27.645.
\end{equation}
This is extremely close to the exact result of -27.703 MeV.

It should be pointed out that the good agreement with the exact
result is mainly due to the fact that, with the HF basis,
the matrix
elements $\langle 1'1'|H'_I|1'2'\rangle_{0,1}$ and
$\langle 1'1'|H'_I|1'2'\rangle_{1,0}$ are very small (see Table 3).
As we have mentioned previously, the sum of these two matrix
elements vanishes in the HF basis [Eq.(\ref{abcd})]. We further notice that
the exact ground-state energy (i.e., the energy obtained in the full-space
matrix diagonalization) of the $A$=4 system, $E_4$, does not depend
the magnitude of each of the two matrix elements, which we denote by
$x$: $x=|\langle 1'1'|H'_I|1'2'\rangle_{0,1}|=
|\langle 1'1'|H'_I|1'2'\rangle_{1,0}|$.
However, it is clear that the effective 2-, 3-, and 4-body
contributions to $E_4$ do depend on $x$. In fact, as
shown in Fig.1, the leading-order diagrams for the effective 2-, 3-, and
4-body forces involve $x$.
Therefore, when we increase $x$,
the effective 2-, 3-, and 4-body interactions will all change but
these changes produce no net effect on $E_4$ which remains
the same. In particular, one can show that as $x$ increases, the
effective 2-body interaction becomes more attractive while the
effective 3-body interaction becomes more repulsive.
The smallness of $x$ in the HF basis means that the effective
many-body interactions are minimized and, consequently, the result
using only the effective two-body interaction is in good agreement
with the exact result.

\section{Conclusions}
With the Lee-Suzuki iterative method, it is now feasible to
calculate the exact, energy-independent, effective two-body interaction
for a no-inert-core model space
from the starting-energy-dependent Bruckner G matrix $G^{(2)}(\omega)$.
We have demonstrated through a simple two-level model that
when only the two-body effective interaction is used for
shell-model calculations in a truncated model space, the choice
of the single-particle basis is very important.
An optimal choice of the single-particle basis should be chosen to
minimize the neglected three- and more-body effective interactions.
We have shown that a self-consistent HF basis serves this purpose very
well for the two-level model.
Encouraged by these results, it is now worth examining the more realistic
situation when one must calculate $G^{(2)}(\omega_2)$ in some chosen
basis {\it before} any self-consistency calculation may be attempted.
The present results indicate that one should choose a realistic
single-particle $H_0$ for each $A$ (assuming realistic is close to
self-consistent) in order to minimize the neglected effective
many-body forces.

\section*{Acknowledgments}
We thank the Department of Energy for partial support
during our stay at the
Institute for Nuclear Theory at the University of Washington,
where this work was initiated.
Two of us (B.R.B. and D.C.Z.) also acknowledge
partial support of this work by the National Science Foundation,
Grant No. PHY91-03011. One of us
(J.P.V.) acknowledges partial support by the U.S.
Department of Energy under Grant No. DE-FG02-87ER-40371, Division
of High Energy and Nuclear Physics and partial support from the
Alexander von Humboldt Foundation.

\vspace{0.2in}

\begin{small}

\end{small}

\vspace{0.5in}

\section*{Figure Caption}
{\bf Fig.1}  The leading-order diagrams for the effective
2-, 3-, and 4-body forces.

\end{document}